\documentstyle[12pt,a4]{article}
\begin{document}

\newcommand{\be}{\begin{equation}}
\newcommand{\ee}{\end{equation}}
\newcommand{\ben}{\begin{eqnarray}}
\newcommand{\een}{\end{eqnarray}}
\newcommand{\nn}{\nonumber}
\newcommand{\n}{\label}
\setcounter{page}{1}

\begin{center}
{\bf A Quintessence Scalar Field in Brans--Dicke Theory}
\end{center}

\begin{center}
Narayan Banerjee\footnote{Permanent address: Relativity and Cosmology
Research Centre, Department of Physics, Jadavpur University, 
Calcutta-700032, India} and 
Diego Pav\'{o}n\footnote{Electronic mail address: diego@ulises.uab.es}\\
{\small Departamento de F\'{\i}sica. Facultad de Ciencias. Edificio Cc. \\
Universidad Aut\'onoma de Barcelona. \\
E--08193 Bellaterra (Barcelona). Spain}
\end{center}

\begin{abstract}
It is shown that a  minimally coupled scalar field in Brans--Dicke theory 
yields a non--decelerated expansion for the present universe for open,
flat and closed Friedmann-Robertson-Walker models.  
\end{abstract}


\section{Introduction}
The recent extensive search for a matter field which can give rise to
an accelerated expansion for the universe stems from the
observational data regarding the luminosity--redshift relation of
the type Ia supernovae up to about $z\sim{1}$ \cite{Perlmutter}.
This matter field is called a ``quintessence matter" (Q--matter for short).
The most popular candidate as a Q-matter has so far been a scalar
field having  a potential which generates a sufficeint negative
pressure at the present epoch \cite{Caldwell}. Amongst the scalar fields 
considered as the Q-matter, the tracker field slowly rolling 
down its potential as proposed by Zlatev, Wang and Steinhardt \cite{Zlatev} 
appears to be very promising.  Some exotic matter like the
domain walls or cosmic strings also find themselves amongst the
possible candidates\cite{Bucher}. Unfortunately most of these fields
work only for a spatially flat $(k=0)$ FRW model. Very recently,
Chimento {\it et al.} \cite{Chimento} showed that a combination of
dissipative effects like a bulk viscous stress and a quintessence
scalar field gives an accelerated expansion for an open universe 
$(k=-1)$ as well. This model also provides a solution for the 
coincidence problem as the ratio of the density parameters
corresponding to the normal matter and the quintessence field
asymptotically approaches a constant value. Recently Bertolami and 
Martins \cite{Bertolami} obtained an accelerated expansion for the universe 
in a modified Brans-Dicke (BD) theory by introducing a potential which is a 
function of the Brans-Dicke scalar field itself. \\ \\
The present work investigates the possibility of obtaining a 
non--decelerating $(q \equiv  - a \ddot{a}/(\dot{a})^{2} \leq{0})$ 
expansion for the universe in Brans-Dicke
theory with the help of another scalar field which is minimally
coupled to gravity and serves as the quintessence matter. Brans-Dicke
theory played a major role in the attempts to solve the ``graceful
exit" problem in the inflationary scenario prevailing in the early
universe. It is worthwhile to explore the possibilty of an
application of the theory towards a reasonable explanation of the
late time behaviour of the universe as well.  We find that for a negative
value of the Brans-Dicke parameter $\omega$, the theory leads to an
accelerated expansion for the universe for a spatially flat model.
The possible solutions contain the Bertolami-Martin's solution as well. 
Furthermore, the theory also leads to at least non--decelerating
expansions even for non flat models. It should be noted that we do
not modify Brans-Dicke theory by introducing a self interaction of
the Brans-Dicke's scalar field, but rather look at the possibility of
a non--decelerating solution with the help of a quintessence field
within  the purview of the theory itself.

\section{Field equations and the solutions}
The Brans - Dicke theory of gravity is given by the action
\be
S = {{1}\over{16{\pi}G_{0}}}\int{\sqrt{-g}}[{\phi}R - 
\omega{\phi_{,\alpha}\phi^{,\alpha}\over{\phi}} + L_{m}]d^{4}x ,
\n{a1}
\ee
where $\phi$ is the BD scalar field, $\omega$ is the dimensionless
constant BD parameter and $L_{m}$ is the Lagrangian for all other matter fields.
If we assume the matter field to consist of a perfect fluid and a
scalar field $\psi$ as the quintessence matter, the field equations for a 
Robertson-Walker spacetime are,
\be
3{(\dot{a}^2+k)\over{a^{2}}} = {(\rho_{m}+\rho_{\psi})\over\phi} 
-3{\dot {a} \dot{\phi}\over{a \phi}}
+{\omega\over{2}}{\dot{\phi}^{2}\over{\phi}^{2}},
\n{fe1}
\ee

\be
2{\ddot{a}\over{a}} + {(\dot{a}^{2}+k)\over{a^{2}}} = 
{-(p_{m}+p_{\psi})\over{\phi}} 
-{\omega\over{2}}{\dot{\phi}^{2}\over{\phi^{2}}} 
-2{\dot{a}\dot{\phi}\over{a\phi}} - {\ddot{\phi}\over{\phi}}.
\n{fe2}
\ee
Here, $a$ is the scale factor of the Robertson--Walker metric, 
$k$ the spatial curvature index, $\rho_{m}$ and  $p_{m}$ are the
density and the pressure of the normal matter, $\rho_{\psi}$ and
$p_{\psi}$ are those due to the quintessence field given by

\be
\rho_{\psi} = \frac{1}{2}{\dot{\psi}}^{2} + V(\psi), \qquad
p_{\psi} = \frac{1}{2}{\dot{\psi}}^{2} - V(\psi),
\n{qm}
\ee
\noindent
where $V=V(\psi)$ is the relevant potential. The wave equation for the
scalar field $\psi$ reads

\be
\ddot{\psi} + 3\frac{\dot{a}}{a} \dot{\psi} = - \frac{dV(\psi)}{d\psi},
\n{we1}
\ee
and the wave equation for the Brans-Dicke scalar field $\phi$ is

\be
{\ddot{\phi}} + {3\dot{a}\dot{\phi}\over{a}}
=\frac{1}{2 \omega +3}[({\rho_{m}} - 3p_{m}) +({\rho}_{\psi}-3p_{\psi})].
\n{we2}
\ee

The matter conservation equation,

\be
\dot{\rho_{m}} + 3{\dot{a}\over{a}}({\rho_{m}} + p_{m}) = 0,
\n{mce1}
\ee
follows from the field equations. Assuming that at the present epoch
the universe is filled with cold matter with negligible pressure, we
put $p_{m}=0$, and this equation  integrates to

\be
\rho_{m} = \rho_{1}/{a^{3}},
\n{mc2}
\ee
where $\rho_{1}$ is an integration constant. Since our principal interest
is to find an accelerating power law solution for the scale factor
$a$, we shall assume that both $a$ and $\phi$ are
power functions of the cosmic time $t$ in the form

\be
a = a_{1}t^{\alpha}, \qquad \phi = \phi_{1} t^{\beta}.
\n{as1}
\ee
We shall look at the possibilities of consistent solutions with $a_{1}, 
\phi_{1}, \alpha, \beta$ are constants 
with $a_{1}, \phi_{1}$ being positve definite and $\alpha \geq{1}$.
These constants will be related amongst themselves and the characteristic 
constants of the theory through the field equations. \\ \\
By combining (~\ref{fe1}) and (~\ref{fe2}) with (\ref{as1})
yields the expression for $\dot{\psi}^{2}$ as

\be
{\dot{\psi}}^{2} = \frac{2k\phi_{1}}{a_{1}^{2}}t^{\beta - 2\alpha}
+(2\alpha + \alpha\beta - \omega\beta^{2} - \beta^{2} +\beta)\phi_{1}
t^{\beta - 2} - \frac{\rho_{1}}{a_{1}^{3}} t^{-3\alpha}.
\n{qm3}
\ee

The potential $V$ can be found out from the equation (~\ref{we2}) as 

\[
V = -\frac{1}{2}\frac{\rho_{1}}{a_{1}^{3}} t^{-3\alpha} + 
\frac{1}{2}\frac{k\phi_{1}}{a_{1}^{2}}t^{\beta - 2\alpha} +
\frac{1}{4} \phi_{1}[(2\omega + 3)(\beta + 3\alpha - 1)\beta
\]
\be
 + (2\alpha + \alpha\beta - \omega\beta^{2} - \beta^{2} +\beta)]t^{\beta
-2}.
\n{pot1}
\ee

The wave equation for the quintessence scalar field $\psi$
(~\ref{we1}), when multiplied by $\dot{\psi}$, looks like 

\be
- \frac{dV}{dt} = \dot{\psi}\ddot{\psi} + 3\frac{\dot{a}}{a}\dot{\psi}^{2},
{\nn}
\ee

which readily yields a first integral and hence an expression for $V$ as 

\[
V =  -\frac{1}{2}\frac{\rho_{1}}{a_{1}^{3}} t^{-3\alpha} - 
\frac{k\phi_{1}}{a_{1}^{2}}\frac{(\beta+4\alpha)}{(\beta-2\alpha)} 
t^{\beta - 2\alpha}
\]
\be 
- (2\alpha + \alpha\beta - \omega\beta^{2} - \beta^{2} +\beta)
\frac{(\beta-2+6\alpha)}{2(\beta-2)}\phi_{1}t^{\beta-2}.
\n{pot2}
\ee

Now we demand that the right hand sides of  (~\ref{pot1}) and (~\ref{pot2}) 
coincide. This leads to the required consistency
relations amongst the constants. These relations lead to quite a few 
possibilities of solutions for an accelerating universe 
($\alpha \geq{1}$). These possibilities fall into two broad
classes, one in which $k=0$ and the second where $k$ is different
from zero.  \\ \\
{\bf{Case 1 : $k=0$}}  \\ 
In this case, the consistency condition is 

\be
6(4\alpha^{2} - 2\alpha + 2\alpha^{2}\beta - \alpha\beta) = 
\omega\beta(\beta^{2} + 6\alpha\beta + 12\alpha - 4).
\n{cc1}
\ee

It is easily seen that the condition (~\ref{cc1}) is automatically
satisfied if $\beta = -2$, and so equation (~\ref{qm3}) reduces to

\be
\dot{\psi}^{2} = -\frac{2(2\omega+3)\phi_{1}}{t^{4}} - \frac{\rho_{1}}
{a_{1}^{3}}t^{-3\alpha}.
\n{qm4}
\ee
This indicates that $\omega < -3/2$ as $\dot{\psi}^{2}$ cannot
be negative. For $\alpha = 4/3$, one has 
$2|2\omega+3|\phi_{1} \geq\frac{{\rho_{1}}}{{a_{1}^{3}}}$. In this case 
equation (~\ref{qm4}) integrates to

\be
\psi = \pm{\frac{A}{t}},
\n{qm5}
\ee
where $A^{2}=-2(2\omega+3)\phi_{1} - \frac{\rho_{1}}{a_{1}^{3}} >0$,  
and a simple form of $V(\psi)$ can be obtained,

\be
V =V_{1} \psi^{4},
\n{pot3}
\ee
$V_{1}$ being a constant, related to the other constants of
integrations like $a_{1}$, $\rho_{1}$ etc. through the field equations.
With $\alpha=4/3$, deceleration parameter $q=-1/4$ and the model
works for all time $0< t < \infty$ provided the condition 
$2 |2\omega+3| \phi_{1} \geq \rho_{1}/a_{1}^{3}$ is satisfied. The
value of $\omega$ is related to the other constants through the relation
\[
2 \omega + 3 = - \frac{\rho_{1}}{a_{1}^{3}\phi_{1}}\pm \sqrt{
\frac{2a_{1}^{2}\phi_{1} -3\rho_{1}}{6a_{1}V_{1}\phi_{1}^{2}}}.
\]

For other values of $\alpha$ the model does not work for the whole range of 
time $0<t<\infty$. If $\alpha > 4/3$, when the rate of acceleration is
faster than $q=-1/4$, the model works only for $t>[\frac{\rho_{1}}
{2a_{1}^{3}\phi_{1}|2\omega + 3|}]^{\frac{1}{3\alpha -4}}$. For 
$1<\alpha<4/3$, the model is valid only up to a time $t = [\frac{\rho_{1}}
{2a_{1}^{3}\phi_{1}|2\omega + 3|}]^{\frac{1}{3\alpha -4}}$ during
which the universe expands with an acceleration with a rate less than 
$q=-1/4$.  \\ \\
For $q=-1/4$, where the model works for the entire time span, the 
present age of the universe can be calculated from (~\ref{fe1}) as 
\be
t_{0}=2\sqrt{2}\left[ \frac{(2\omega+3)-\frac{\rho_{0}}{2a_{0}^{3}\phi_{0}}-
V_{1}A^{4}}{3(3\omega+4)}\right] ^{1/2}\frac{1}{H_{0}}.
\nn
\ee
In the large $\omega$ limit, this equation reduces to
\be
t_{0}\simeq{\frac{8 \sqrt{- 2\, \omega V_{1}}}{3H_{0}}},
\nn
\ee
where $\omega$ is obviously a negative quantity. Choosing 
$V_{1} = \textstyle{- 9\over{128}} \omega^{-1}$ and 
$H_{0} \simeq 65$ km$\cdot$s$^{-1}/$Mpc, the present 
age of the Universe turns out to be approximately $15$ Gyr. \\ \\
Likewise, the present rate of variation of $G$ is $|\dot{G}/G|_{0}
=|\dot{\phi}/\phi|_{0}= \textstyle{3\over{2}}H_{0}<10^{-10}$ per year.
This is quite compatible with the observational data 
(see \cite{Weinberg} and references therein).\\  \\
All these possibilities are for $\beta = -2$. For other values of
$\beta$ also, one can obtain consistent solutions. Such as if $\beta
= -1$, the equation (~\ref{cc1}) yields two possible solutions for
$\alpha$, namely $1/2$ and $-\omega/2$.  We disregard $\alpha = 1/2$
as we are interested in non--decelerating models where
$\alpha \geq{1}$. For $\omega=-2$, we have $\alpha = 1$, i.e. an
uniformly expanding universe with $q=0$. In this case, $\phi_{1}$
should be greater than $\rho_{1}/a_{1}^{3}$. The equation
system can be easily solved to get $\psi \propto {t^{-1/2}}$ and 
$V \propto {\psi^{6}}$. In this uniformly expanding scenario, the model
can work for the whole range of $0<t<\infty$. If $\omega$ is further 
negative, i.e., $\omega\leq{-2}$, we get faster rates of acceleration,
but the model does not work for the whole range of time. \\ \\
{\bf{Case 2 : $k\neq{0}$}} \\
In this case the condition (~\ref{cc1}) remains in place and a
further condition from (~\ref{pot1}) and (~\ref{pot2}) is 

\be
\beta = -2\alpha.
\n{cc2}
\ee

It can be seen from the field equations (~\ref{fe1}) and (~\ref{fe2})
that one can obtain consistent solutions only for $\alpha = 1$. So equation 
(~\ref{cc2}) immediately yields $\beta = -2$. With this value of $\beta$, 
the condition (~\ref{cc1}) is automatically satisfied for all values of 
$\omega$. In this case also the model works for a  limited
period of time, $0<t<t_{1}$, where

\be
t_{1} =2\phi_{1}\left[\frac{k}{a_{1}^{2}}-(2\omega+3)\right]
\frac{a_{1}^{3}}{\rho_{1}}.
{\nn}
\ee

For an open universe, i.e. for $k=-1$, $(2\omega+3)$ has to be
negative and $ |2\omega+3|>1/a_{1}^{2}$. \\ 
For a closed universe $(k=1)$, the model works even for a positive 
$(2\omega+3)$ provided $(2\omega+3) a_{1}^{2}< 1$. If however, 
$(2\omega+3)$  is negative, the model holds good without any such 
relation between $a_{1}$ and $\omega$. \\
Thus, unlike most of proposed models with a non--positive definite 
deceleration parameter, a quintessence field in Brans-Dicke theory works 
for a spatially non--flat Robertson Walker spacetime as well. It is 
true that for $k\neq{0}$ cases the only consistent solutions have $q=0$,
but it is anyway non--decelerating and thus the expansion rate is
faster than $t^{2/3}$, which might sufficiently explain the recent
observations on the distant supernovae \cite{Riess}. \\ \\
It is worthy of note that the field equations can be integrated to
produce consistent solutions for other forms of normal matter. Such
as for a radiation dominated universe $( p_{m}= \rho_{m}/3)$, a
simple choice of $V$ as $V=V_{0}\psi^{6}$ and a negative $\omega$ will 
lead to a decelerated expansion $(q>0)$ for the universe with  
$a\propto{t^{\frac{3}{4}}}$  and $\phi\propto{t^{-1}}$. Thus the model 
can be interpolated back to earlier epoch to yield a decelerated universe 
which is required in order to explain processes like nucleosynthesis.

\section{A possible solution to the flatness problem}
One important aspect of this model is that potentially it can solve
the flatness problem as well. To see this we effect a conformal
transformation  as

\be
\bar{g}_{\mu\nu} = \phi g_{\mu\nu} ,
\nn
\ee
which enables us to identify the energy contributions from different
components of matter very clearly. 
Equation (~\ref{fe1}), in this new version, looks like

\be
{3(\dot{\bar{a}}^{2} + k)\over{\bar{a}^{2}}} = \bar{\rho}_{m} + 
\bar{\rho}_{\phi} + \bar{\rho}_{\psi},
\n{nf1}
\ee
where a bar indicates quantities in the transformed version;
$\rho_{m}$, $\rho_{\phi}$, $\rho_{\psi}$ are the contributions to the
energy density from the normal matter, the BD scalar field and the 
quintessence scalar field respectively. The quantity $\bar{\rho}_{\phi}$ 
is actually given by

\be
\bar{\rho}_{\phi} = \frac{(2\omega+3)}{4}\left(\frac{\dot{\phi}}{\phi}
\right)^{2} = \bar{p}_{\phi}.
\n{ebd}
\ee

The quantities describing the matter distribution are transformed as
$\bar{\rho}_{i}=\phi^{-2}\rho_{i}$ and $\bar{p}_{i}=\phi^{-2}p_{i}$. 
Now we define the dimensionless density parameter $\bar{\Omega}$ as 

\be
\bar{\Omega} = \frac{\bar{\rho}}{3\bar{H}^{2}} = \bar{\Omega}_{m} +
\bar{\Omega}_{\phi} + \bar{\Omega}_{\psi},
\n{dp1}
\ee
where the total density $\bar{\rho}=\bar{\rho}_{m}
+ \bar{\rho}_{\phi} + \bar{\rho}_{\psi} $, and the individual density
paramaters $\bar{\Omega}_{i}$ are defined accordingly. Using the
equation (~\ref{nf1}) and the equation for the conservation for the
total energy,

\be
\dot{\bar{\rho}}+3\dot{\bar{H}}(\rho +p) = 0,
\n{mce2}
\ee

\noindent
we can write down the evolution equation for the density parameter as

\be
\dot{\bar{\Omega}} (\bar{\Omega}-1)(3\gamma - 2)\bar{H}=0.
\n{dp2}
\ee.

The net barotropic index $\gamma$ is defined as 

\be
\gamma\bar{\Omega}=\gamma_{m}\bar{\Omega}_{m} + 
\gamma_{\phi}\bar{\Omega}_{\phi} + \gamma_{\psi}\bar{\Omega}_{\psi}.
\n{dp3}
\ee

The individual $\gamma_{i}'s$ are defined by the equation $ p_{i} =
(\gamma_{i}-1)\rho_{i}$. The ratioes $p_{i}/\rho_{i}$ remain
the same in both  frames and thus $\gamma_{i}$'s do not vary. For our 
choice of matter, $\bar{p}_{m}=0$ and $\bar{p}_{\phi} = \bar{\rho}_{\phi}$, 
and thus $\gamma_{m}=1$ and $\gamma_{\phi}=2$. The third index 
$\gamma_{\psi}$, however, is not a constant and evolves with time via
the equation $\gamma_{\psi}= (p_{\psi}+\rho_{\psi})/(\rho_{\psi}) = 
\dot{\psi}^{2}/(\textstyle{1\over{2}}\dot{\psi}^{2} + V)$.\\
Equation (~\ref{dp2}) indicates that $\bar{\Omega}=1$ is indeed a
solution. The stability of this solution demands that 
$(\partial{\dot{\bar{\Omega}}}/\partial{\bar{\Omega}})_{H}$ should be
negative at $\bar{\Omega} = 1$.  Equation (~\ref{dp2}) shows that for
an expanding universe ($\bar{H}>0$) this is possible only if 
$\gamma<2/3$. From equation (~\ref{dp3}) it can be shown that the relevant
condition for $\gamma<2/3$ is 

\be
\bar{\Omega}_{m} + 4\bar{\Omega}_{\phi} < (2-3\gamma_{\psi})
\bar{\Omega}_{\psi},
\nn
\ee
which can be achieved by a suitable adjustment of the parameters.\\ \\
It is wellknown that geodesic equations are not valid in this
conformally transformed version \cite{Dicke} and hence different
quantities are not dependable regarding the content of their physical
meaning. But it must be emphasized that the character of $k$ remains
unaltered, and thus if $\Omega_{k} = k/a^{2}$ is zero in one
frame, it must be so in the other  as well.

\section{Concluding remarks}
A quintessence scalar field in Brans-Dicke theory is shown to give
rise to an accelerated expansion for the present universe. Bertolami and
Martins \cite{Bertolami} modified Brans-Dicke theory by introducing a
potential function $V=V(\phi)$ where $\phi$ is the Brans-Dicke scalar
field. As Brans-Dicke theory by itself is, in a sense, self--interacting 
(the kinetic term in the action contains $\phi$), we do not include such 
a potential. Rather a  non--gravitational field $\psi$ with a potential 
$V=V(\psi)$ is included. For a spatially flat universe, the model yields 
various non--decelerating solutions including a uniformly expanding 
solution $(q=0)$. For a simple choice of the 
potential $(V \propto \psi^{4})$, the model gives the solution of
\cite{Bertolami}. In this last case, the model can work for all
time $0<t<\infty$. In most of the other accelerating solutions, the
model is seen to work for only a restricted period of time. \\ \\
An important merit of this ansatz is that it provides solutions for a
non flat $(k\neq{0})$ Robertson Walker metric as well. Although these
solutions are not accelerating, they are not decelerating either $(q=0)$.
So along with providing a non--decelerating solution, it can
potentially solve the flatness problem too. In fact it has been shown
that $\Omega=1$ could be a stable solution in this model. \\ \\
This acceleration of the universe is achieved, in general, by a
negative $\omega$. It has been claimed that the value of $\omega$
should be large $(>500)$ if Brans-Dicke theory has to be consistent with 
the astronomical observations \cite{Will}. But this value
actually refers to the magnitude of $\omega$ which can still be very
large in this model except in some cases, such as for  $(k=0, q=0)$, 
where $\omega=-2$. Furthermore, reconciliation with Kaluza-Klein theory or 
low--energy string theory favours a negative value of $\omega$ 
(see e.g. \cite{kolitch}).

\section*{Acknowledgements} 
This work has been partially supported by Spanish Ministry of Education 
under Grant PB94--0781. One of us (NB) is grateful to the ``Direcci\'o 
General de la Recerca" of the Catalonian Government for financial 
support under grant PIV99. The authors are grateful to Winfried 
Zimdahl for critically reading an earlier draft of this paper and 
useful suggestions.


\begin{thebibliography}{99}
\bibitem{Perlmutter} 
Perlmutter S {\it et al.} 1999 {\em Astrophys. J.} {\bf 517} 565;
Riess AG {\it et al.} 1998 {\em Astron. J.} {\bf 116} 1009;
Garnavich PM {\it et al.} 1998 {\em Astrophys. J} {\bf 509} 74
\bibitem{Caldwell}
Caldwell RR, Dave R and  Steinhardt PJ 1998 {\em Phys. Rev. Lett.} 
{\bf{80}} 1582; Turner MS and White M 1997 {\em Phys. Rev. D} {\bf 56}
R4439
\bibitem{Zlatev}
Zlatev I,  Wang L and Steinhardt PJ 1999 {\em Phys. Rev. Lett.}
{\bf 82} 896; P.J. Steinhardt, L. Wang and I. Zlatev 1999
{\em Phys. Rev. D} {\bf 59}, 123504 
\bibitem{Bucher}
Bucher M and Spergel D 1999 {\em Phys. Rev. D} {\bf 60} 043505;
Bucher M 1999 [astro-ph/9908047]
\bibitem{Chimento}
Chimento LP, Jakubi AS and Pav\'{o}n D 2000 {\em Phys Rev. D} (in the
press) [astro-ph/0005070]
\bibitem{Bertolami}
Bertolami O and Martins PJ 2000 {\em Phys. Rev. D} {\bf 61} 064007
\bibitem{Weinberg}
Weinberg S 1972 {\it Gravitation and Cosmology} (New York: Wiley)
\bibitem{Riess}
Riess AG 2000 ``The case for an accelerating universe from supernovae"
[astro-ph/0005229]. 
\bibitem{Dicke}
Dicke RH 1962 {\em Phys. Rev.} {\bf 125} 2163
\bibitem{Will}
Will CM 1993 {\it Theory and Experiment in Gravitational Physics}
(Cambridge: Cambridge University Press, 3rd edition)
\bibitem{kolitch}
Kolitch SJ and Eardley DM 1995 {\em Ann. Phys. (N.Y.)} {\bf 241} 128 
\end{thebibliography}
\end{document}